\documentclass[conference]{IEEEtran}
\usepackage{cite}
\usepackage{amsmath,amssymb,amsfonts}
\usepackage{algorithmic}
\usepackage{amsmath}
\usepackage{graphicx}
\usepackage{textcomp}
\usepackage{xcolor}
\usepackage{makecell}
\usepackage{url}
\usepackage{dblfloatfix} 
\usepackage{subfigure}
\usepackage{balance}

\usepackage[left=0.75in,right=0.75in,top=0.75in,bottom=0.75in]{geometry}

\begin{document}
% \title{Humanoid Robot Affects Human Rationality: A Case Study in a Competitive Game}

% \title{Effects of A Robot's Expressive Language on Strategy and Perceptions in a Competitive Game}

\title{\vspace{0.25in}
A Robot's Expressive Language Affects Human Strategy and Perceptions in a Competitive Game}

\author{
\IEEEauthorblockN{Aaron M. Roth$^1$\quad Samantha Reig$^2$\quad Umang Bhatt$^3$\quad Jonathan Shulgach$^3$\quad \\ Tamara Amin$^3$\quad Afsaneh Doryab$^2$\quad Fei Fang$^4$\quad Manuela Veloso$^5$\quad\\
% \affaddr{$^1$Robotics Institute}
% \affaddr{$^2$Human-Computer Interaction Institute} \\
% \affaddr{$^3$College of Engineering} 
% \affaddr{$^4$Institute for Software Research}
% \affaddr{$^5$Machine Learning Department} \\
% \affaddr{Carnegie Mellon University} \\
% \affaddr{Pittsburgh, PA, USA} \\
$^1$Robotics Institute
$^2$Human-Computer Interaction Institute \\
$^3$College of Engineering 
$^4$Institute for Software Research
$^5$Machine Learning Department \\
Carnegie Mellon University \\
Pittsburgh, PA, USA \\
% \email{Corresponding author email: amroth@umd.edu} \\
Corresponding author email: amroth@umd.edu \\
}
}

%\author{\IEEEauthorblockN{Aaron M. Roth\IEEEauthorrefmark{1},
%Samantha Reig\IEEEauthorrefmark{2},
%Umang Bhatt\IEEEauthorrefmark{3},
%Jonathan Shulgach\IEEEauthorrefmark{3}, \\ 
%Tamara Amin\IEEEauthorrefmark{3},
%Afsaneh Doryab\IEEEauthorrefmark{2}, 
%Fei Fang\IEEEauthorrefmark{5}
%, and
%Manuela Veloso\IEEEauthorrefmark{6}
%}
%\IEEEauthorblockA{\IEEEauthorrefmark{1}Robotics Institute, %\IEEEauthorrefmark{2}Human-Computer Interaction Institute %\IEEEauthorrefmark{4}
%, \IEEEauthorrefmark{5}Institute for Software Research  \\ %IEEEauthorrefmark{6}Machine Learning Department,
%\IEEEauthorrefmark{4}
%\IEEEauthorrefmark{3}College of Engineering \\ %\IEEEauthorrefmark{4}School of Computer Science \\ 
%Carnegie Mellon University,
%Pittsburgh, PA\\ Contact email: aaronr1@andrew.cmu.edu }

%}

% \IEEEauthorblockA{\IEEEauthorrefmark{2}Twentieth Century Fox, Springfield, USA\\
% Email: homer@thesimpsons.com}
% \IEEEauthorblockA{\IEEEauthorrefmark{3}Starfleet Academy, San Francisco, California 96678-2391\\
% Telephone: (800) 555--1212, Fax: (888) 555--1212}
% \IEEEauthorblockA{\IEEEauthorrefmark{4}Tyrell Inc., 123 Replicant Street, Los Angeles, California 90210--4321}


\maketitle

\begin{abstract}
As robots are increasingly endowed with social and communicative capabilities, they will interact with humans in more settings, both collaborative and competitive.  We explore human-robot relationships in the context of a competitive Stackelberg Security Game.  We vary humanoid robot expressive language (in the form of ``encouraging'' or ``discouraging''  verbal commentary) and measure the impact on participants' rationality, strategy prioritization, mood, and perceptions of the robot. We learn that a robot opponent that makes discouraging comments causes a human to play a game less rationally and to perceive the robot more negatively.  %We also discover that humans may consider an opponent robot to be a ``distraction'' regardless of the its affect. 
% We also contribute a Natural Language Processing framework to generate expressive comments for a social robot.
We also contribute a simple open source Natural Language Processing framework for generating expressive sentences, which was used to generate the speech of our autonomous social robot.% sentences for our autonomous social robot to say during the study.
\end{abstract}
%-%Afsaneh comments:
%State the contributions in the intro
%- Add the main results to the captions in Figure 3. People want to quickly understand what the graphs show rather than finding it in the text.
%- Looks like you consider the main contribution to be the findings and not the methodology. Haven't you developed a customized affect-aware NLP model and showed it worked? This is a contribution. Also, I don't know how novel the quantal response model is and if you have added anything new to it. Fei can better tell. But if you have, please state it as a contribution too. And in the discussion, address the generalizability of you NLP and QR models.  

% \fi
\IEEEpeerreviewmaketitle

\section{Introduction}
The future will bring humans into contact with robots in a variety of unstructured interactions, many of which will involve engaging robots in verbal dialogue. This includes in-store sales ~\cite{gross2009toomas}, education~\cite{hyun2010relationships}, service interactions~\cite{de2018towards}, and rehabilitation~\cite{broadbent2010attitudes}. In any interaction like this, linguistic nuances and positive or negative valence of social behavior will impact the result of the interaction. In some of these settings, one can imagine a robot and human may have different or even conflicting goals. For example, in a sales setting, a robot completing a sale may prioritize convincing a customer to buy a product, whereas the customer aims to make the optimal decision to satisfy their needs. The humans and the robots have to behave strategically in such settings to gain advantage in the interaction.
%In a sales setting, a robot completing a sale may prioritize convincing a customer to buy a product, whereas the customer aims to prioritize satisfying their needs. Linguistic nuances and positive or negative valence of social behavior would be predominant in both parties reaching an accord within these types of interactions.
Much work has gone into understanding how humans and social robots interact and partner in cooperative settings \cite{affectjoint, cooperate, xu2014robot}. For example, positive robot affect has been shown to contribute to perceptions of robots as teammates~\cite{castellano2009s}.
However, less research has been done to understand how affect impacts interactions when the interests of the humans and robots are not perfectly aligned. 
In this study, we focus on a competitive setting and study the impact of a robot's affect on humans' \textit{rational} behavior, which is understudied in HRI despite its significance. Acknowledging that \textit{affect} takes many forms, we focus on affect exhibited through \textbf{encouraging} and \textbf{discouraging} language. This \textbf{expressive language} is one manifestation of how positive/negative affects could emerge in a competitive interaction.

In this study, we examine the impact of expressive language from a robot on human rationality and strategy prioritization in a representative general-sum competitive game, the Guards and Treasures game. (This paper considers rationality in the context of maximizing expected utility.)  % The literature on Stackelberg security games mostly pertains to human play data, which is collected and analyzed to understand the bounded rational behavior of humans
This game has been used extensively in the literature on Stackelberg Security Games to collect human play data and analyze bounded rationality of humans~\cite{pita2009effective, pita2012robust}. 
We adopt this game as it provides a simple environment in which players' interests are not fully aligned.
We seek to answer the question: how does %the %perceived
encouraging or discouraging language from a humanoid robot opponent impact a human's rationality and strategy in %a game theoretic setting?  %We term the ``perceived encouragement/discouragement'' as ``motivational affect''
this example strategic game? We expect the results from this study to shed light on more general settings of competitive or semi-competitive interaction between robots and humans.
We implement a system to play the game autonomously with dialogue generated by our expressive language algorithm. In a between-subjects study, 40 participants played the game with a humanoid robot. Each participant was exposed to one of two conditions in which the robot made either encouraging or discouraging comments. We analyze the collected data to obtain insights into how the robot's behavior impacts participants' rationality and emotions. 
Some existing work shows that threatening behavior from a robot may increase humans' attentional control~\cite{Spatolaeaat5843}. In contrast, in our study, discouraging comments from a robot decreased a participant’s rationality during gameplay. In addition, negative language contributed to negative social attributions to the robot. 

In addition to investigating the impact of a robot's use of expressive language on a human opponent's strategy, risk-taking, and performance in a competitive game setting, we contribute an open-source Natural Language Processing (NLP) model that is affect-aware.
% To our knowledge, no-one has investigated the impact of a robot's use of expressive language on a human opponent's strategy, risk-taking and performance in a competitive game settings. We contribute methodology to conduct such an investigation including developing a customized, open-source NLP model that is affect-aware, and extending the work of \cite{yang2011improving} wherein the best response quantal response is used to identify the rationality of a player in a Stackelberg Security game. We contribute findings that show that a discouraging robot leads to lower rationality while an encouraging robot is associated  with  higher  rationality. 
We discuss these findings and others in section \ref{sec:results}.

\section{Background and Related Work}\label{sec:rw}

%Humans interpret gestures and verbal cues which contain information about their mental and emotional states~\cite{frijda2005emotion}. 
Observing others' moods can have specific consequences for the observer \cite{wild2001emotions}, %and can 
impacting their performance~\cite{conmy2008trash,Spatolaeaat5843}, risk taking~\cite{steinberg2008social}, decision making~\cite{russell2003core}, and mood~\cite{barsade2002ripple}. Mood-contagion is a well-researched automatic mechanism whereby the observation of another person's emotional expression induces a congruent state of mood in the observer. 

% \newgeometry{left=0.75in,right=0.75in,top=0.75in,bottom=0.75in}

Affect is a general term relating to emotions, moods, feelings and desires that may influence behavior. Affective states vary in their degree of \textit{activation} (intensity) and \textit{valence} (whether they are positive and negative) \cite{22}. 
Research has shown that humans' perception of robots' affect and/or expressive language can influence interactions.  Various studies have used the ROMAN robot for facial expressions\cite{55,56}, NAO for body expressions\cite{55,56,57}, KOBIAN for body and facial expressions\cite{54}, and Cozmo for nonverbal behaviors\cite{skaageby2018well}. One study found that humans can identify and respond to a robot's expressive language~\cite{46}. Robot language can influence the effectiveness of assistive tasks including learning or receiving vocal encouragement expressions~\cite{moody}. Another study found that a computer agent's expression (anger and happiness) impacted the way humans negotiated with it~\cite{anger}.
Research on the impact of expressive language between language models in human-robot interaction has been limited to joint human-robot tasks in a cooperative setting; therefore we extend an expressive language model to a competitive human-robot interaction situation.

One experiment showed that human strategy is different when facing a text-based mediator vs. a mediator with an avatar~\cite{lin2011bridging}, which indicates that the form of interaction between a robot and a human matters. Several other studies have shown the impact of affective virtual agent behavior on human task performance \cite{x106, xu2014robot, leite2008emotional}. We consider a competitive scenario and analyze human performance and strategy when facing a humanoid robot.
% We consider to what degree this holds true for a humanoid robot

To our knowledge, we are the first to explore the impact of a robot's expressive language statements on human performance in a competitive game setting.  We use a %competitive Stackelberg security game
game which can be mathematically modeled as a Stackelberg security game~\cite{yang2011improving} to analyze human rationality and strategy %via quantal response (participant rationality) 
as well as human perception of the robot from a combination of validated scales and experiment-specific Likert items. %Stackelberg security games are a common choice to understand human rationality because they contain an optimal strategy and have the ability to measure quantal response and subjective utility quantal response (participant strategic prioritization). 
Quantal response~\cite{mckelvey1995quantal} and its variant---subjective utility quantal response~\cite{nguyen2013analyzing}---have been proposed and used in game theory literature to quantitatively model the bounded rationality of human players and their prioritization of different factors impacting their decision making.  We leverage these models for strategic human-robot-interaction.
We use a humanoid robot in this study. In contrast to \cite{xu2014robot} and \cite{leite2008emotional}, we deal with verbal affect cues instead of gesture and posture.

\section{Methodology and Contributions}\label{sec:methods}

\subsection{Overview of Study}\label{sec:methods_overview}

The primary goal of our study was to determine the effect of a robotic opponent's expressive language
%affective state
on a human's game-playing strategy in a %game theoretic 
competitive and strategic game. %setting. 
We hypothesized that (1) when playing a competitive game against a humanoid robot, a human's strategy will be influenced by the expressive language of the robot, and (2) encouraging expressive language will positively impact participants' social perceptions of the robot. We conducted a between-subjects experiment in which a human played a repeated Stackelberg security game against a robot. (Specifically, the human plays two practice rounds without the robot and then 35 rounds of the game ``against'' the robot.) The robot made either \textit{encouraging} or \textit{discouraging} comments during game play. We recorded participants' actions to understand the nature of their game play strategy. We also measured their perceptions of the task, of their performance, and of the robot. A detailed description of the experimental setup and procedure can be found in section \ref{sec:experiment}.

We manipulated robot expressive language in the form of periodic utterances that were either \textit{encouraging} or \textit{discouraging}.  Utterances were generated via an NLP model we discuss in section \ref{sec:methods_NLP}.  The robot exhibited optimal strategy in all games, regardless of the condition or the human player's actions.  The robot moved and spoke autonomously according to a script that our framework generated ahead of time.% before each study session. (i.e., it did not respond to participants' actual comments and game moves, but instead gave the illusion that it did).  
%according to a randomized drawing from a script that combined human-sourced and robot generated speech.  human templates and robot generated.
%   The robot moved and spoke autonomously according to a predefined script (i.e., it did not respond to participants' actual comments and game moves, but instead gave the illusion that it did).  

Our primary measures of interest pertained to the participant's strategy.  We analyzed their ``strategy'' in two ways. First, we used a quantal response equilibrium model to determine the degree to which the human played rationally.  Second, we evaluated the nature of the strategy itself, in terms of which aspects of the game environment the participant prioritized in their decision-making process (assessed via the parameter values of the subjective utility quantal response model that can best fit the human play data and via self-report).  Both of these are discussed in greater detail in section \ref{sec:methods_QR}.
Other variables of interest were social perceptions, mood, and perceived robot mood, measured in terms of 1) participants' answers to questions about themselves and the robot along Likert scales, and 2) answers to free-response questions about perceptions of the robot and the task after the game is played.

\begin{figure}[t]
%\centerline{}
\centering
\includegraphics[width=0.5\columnwidth]{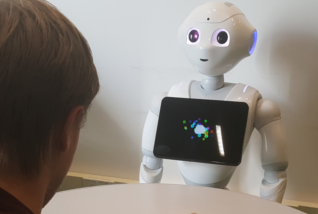}
\caption{The study setup from a participant's perspective}
\label{fig:study}
\end{figure}

\subsection{NLP Model}\label{sec:methods_NLP}

To give our robot expressive language, we developed an affect-aware bidirectional fill-in-the-blank N-Gram model.
%An N-gram model trains on corpora and counts how often each word follows each preceding N-words. From these counts, we construct a probability that any particular word follows a previous sequence of words.
% This probability is derived by computing the frequency of that particular word's occurrence after a given sequence compared to the frequency of any observed word following the same sequence.
% After creating counts of sequences of words of length N+1, the probability can be expressed as:

We construct a probability that a particular word follows a previous sequence of words using an N-gram~\cite{brown1992class} method:

% \begin{equation}\label{eqn:ngram}
% \begin{split}
% P(w_n | & w_{n-(N-1)}, ..., w_{n-1}) = \\
%  & \frac{\text{Count}(w_{n-(N-1)},...,w_{n-1},w_n) + \alpha} { \text{Count}(w_{n-(N-1)},...,w_{n-1},\text{*}w) + D\alpha }
% \end{split}
% \end{equation}

\begin{equation}\label{eqn:ngram}
\begin{split}
P(w_n | & w_{S}, ..., w_{n-1}) = \frac{\text{C}(w_{S},...,w_{n-1},w_n) + \alpha} { \text{C}(w_{S},...,w_{n-1},\text{*}w) + D\alpha }
\end{split}
\end{equation}

where $S = n-(N-1)$, C() means ``count of'', and $\text{*}w$ is a wildcard meaning ``any word observed as completing this sequence''. Thus, $P(w_n | w_{S}, ..., w_{n-1})$ is the probability that a particular word $w_n$ follows a particular sequence of $N$ other words.  In our usage, as shown, we add $+ \alpha$ and $+ D\alpha$ terms as Laplacian smoothing to account for situations where a word was not observed. We use $\alpha=1$ and $D=$ [number of words that could fit *$w$ for the given preceding sequence].

% To make the language not just natural but connoting a specific emotion (``affect-aware''), we took advantage of the AFINN Affect Dictionary, which rates the emotional valence of a word on a scale from -5 to 5 \cite{nielsen2011new}.  During the game (after the model is trained), we feed in a bank of neutral sentence stems (sentences with fill-in-the-blanks).  The model selects words to fill in the blanks that are appropriate given the sentence and that give the sentence appropriate affect. This also allows us to use bidirectional N-grams--training and predicting based on the words preceding \textit{and} following a word to be predicted. We use both bigrams and trigams (N=2, N=3) in the forward and reverse direction.
To make the language affect-aware, we used the AFINN Affect Dictionary, which rates the emotional valence of a word~\cite{nielsen2011new}. We constructed sentence stems (sentences with fill-in-the-blanks) such that positive or negative fill-in words result in encouraging or discouraging sentences. (These sentence stems can be fed into our model.)
%During the game (after the model is trained), we feed in a bank of neutral sentence stems (sentences with fill-in-the-blanks).  The model selects words to fill in the blanks that are appropriate given the sentence and that give the sentence appropriate affect. This also allows us to use bidirectional N-grams--training and predicting based on the words preceding \textit{and} following a word to be predicted. 
We use both bigrams and trigams ($N=2$, $N=3$), and train our model in both forward and reverse direction.
The final equation used to select the words to complete neutral sentence stems is given by:

\begin{equation}\label{eqn:ngram_final}
\begin{split}
P(w_n | & w_{n+2}, w_{n+1}, w_{n-1}, w_{n-2}) = z_5 * V(w_n) * A \\
&+ z_1 * P(w_n | w_{n+2}, w_{n+1}) + z_2 * P(w_n | w_{n+1}) \\
&+ z_3 * P(w_n | w_{n-2}, w_{n-1}) + z_4 * P(w_n | w_{n-1}) \\
% &+ z_5 * V(w_n) * A \\
 A \in &\{-1,1\}, \quad \sum_{i=1}^{5}{z_i} \leq 1
\end{split}
\end{equation}

where the probabilities on the right-hand side are calculated according to %equation
(\ref{eqn:ngram}), $V$ gives the AFINN affective valence of a word (or 0 if not in the dictionary), $A$ indicates whether the affect is encouraging (+1) or discouraging (-1), and the $z_i$ values are weights. We train our model on transcripts of popular films from the IMSDb archive \cite{walker2011perceived,walker2012annotated}.  %During training and prediction, a set of ``stop words'' (such as ``is'' and ``and'') is ignored for the purposes of defining sequences of words. Numerals and punctuation are also filtered out before training. Additionally, during prediction, certain words are ignored (blacklisted), such as ``kill''.  This is to prevent the robot from being too vulgar, saying things that do not make sense, hoping for a participant's death, or making comments that are otherwise offensive or uninterpretable.
The code to generate the model from any corpora and make predictions based on arbitrary sentence stems can be found on Github.\footnote{Find the NLP model code here: \url{https://github.com/AMR-/fill_in_the_blank_word_prediction}}
% DEL-OPTION: foodnote above

\subsection{Quantal Response}\label{sec:methods_QR}

\subsubsection{Measure of Degree of Rationality}\label{sec:methods_QR_lambda}

Each participant played several rounds of Guards and Treasures, a Stackelberg security game~\cite{yang2011improving,sinha2018stackelberg}, against the robot.  For the purposes of the rationality calculation described here, note that during each round, the participant chose a single action from a set of $N$ options, % based on provided information,
in an attempt to maximize the expected numerical reward.

Quantal response model assumes that a human player is more likely to choose more promising options. Mathematically, let $q_{c_r}$ represent the probability of the participant selecting choice $c$ in round $r$. It is defined in %equation 
\eqref{eqn:QR_Lb}, where $\lambda$ %represents the amount of noise in the participant's response (that is, $\lambda$ captures how rational the participant's decision is). 
is a parameter that can control or represent how rational the participant's decision is.

\begin{equation}\label{eqn:QR_Lb}
\begin{split}
    q_{c_r} = \space  
        \frac{\exp{(\lambda U_{c_r,r})}} {\sum_{j=1}^{N} {  \exp{ ( \lambda U_{j,r} ) }}}
\end{split}
\end{equation}

%For a given set of rounds $\Upsilon$, we fit $\lambda$ to the game play data by estimating the maximum likelihood of the quantal response $q_{c_r}$. 
Assuming that participants follow the quantal response model, given the game play data of a given set of rounds $\Upsilon$, we can learn the value of parameter $\lambda$ that can best fit the data via maximum likelihood estimation.
This is shown in %equation 
\eqref{eqn:QR_La} where $\lambda$ is fit to a subset of rounds $\Upsilon$.

% OLD 3 / 3/4
% \begin{equation}\label{eqn:QR_L}
% \begin{split}
%     \lambda = \space \operatorname*{arg \space max}_{\lambda} \sum_{r \in \Upsilon}\log (q_{c_r}), \quad
%         q_{c_r} = \space  
%         \frac{\exp{(\lambda U_{c_r,r})}} {\sum_{j=1}^{N} {  \exp{ ( \lambda U_{j,r} ) }}}
% \end{split}
% \end{equation}

\begin{equation}\label{eqn:QR_La}
\begin{split}
    \lambda = \space \operatorname*{arg \space max}_{\lambda} \sum_{r \in \Upsilon}\log (q_{c_r})
\end{split}
\end{equation}

where $U_{i,r}$ is the known real utility of choice $i$ in round $r$ (see \eqref{eq:U}), $c_r$ is the number of the choice chosen by the participant in round $r$, and $\Upsilon$ is the subset of rounds to be used in the calculation. %A $\lambda = 0$ would indicate random behavior, while a $\lambda = \infty$ would indicate perfect rationality.  (It is not a linear scale.)  

\subsubsection{Measure of Prioritization in Strategy}\label{sec:methods_QR_strat}
We can also follow the Subjective Utility Quantal Response (SUQR) model, defined in \eqref{eqn:QR_Wa}, to determine the probability $s_{c_r}$ that a participant selects choice $c$ in round $r$ based on the parameters $W$ representing the \textbf{strategic priority} of the participant \cite{nguyen2013analyzing}.  $W$ denotes the importance to the participant %reward, penalty, and probability of guard present are to the participant.
of different attributes of each of the options.

\begin{equation}\label{eqn:QR_Wa}
\begin{split}
    s_{c_r} = \frac {\exp{(W^{T} X_{c_r,r})}} {\sum_{j=1}^{8} {  \exp{ ( W^{T} X_{j,r} ) }  } } \\
W^{T} = [w_1 \quad w_2 \quad ... \quad w_n] & \qquad X_{i,r}^{T} \in \mathcal{R}^3
\end{split}
\end{equation}
where $X$ represents a vector of values for attributes of the choices, and $W$ is the strategic prioritization showing how much weight a participant gives to each attribute.

% \begin{equation}\label{eqn:QR_W}
% \begin{split}
%     W = \space \operatorname*{arg \space max}_{W} \sum_{r \in \Upsilon}\log (s_{c_r}),& \quad s_{c_r} = \frac {\exp{(W^{T} X_{c_r,r})}} {\sum_{j=1}^{8} {  \exp{ ( W^{T} X_{j,r} ) }  } } \\
% W^{T} = [w_1 \quad w_2 \quad ... \quad w_n] & \qquad X_{i,r}^{T} \in \mathcal{R}^3
% \end{split}
% \end{equation}
% old 4 / 5/6
% \begin{equation}\label{eqn:QR_W}
% \begin{split}
%     W= \operatorname*{arg \space max}_{W} \sum_{r \in \Upsilon}\log (s_{c_r}),\quad & s_{c_r} = \frac {\exp{(W^{T} X_{c_r,r})}} {\sum_{j=1}^{8} {  \exp{ ( W^{T} X_{j,r} ) }  } } \\
% W^{T} = [w_1 \quad w_2 \quad ... \quad w_n] & \qquad X_{i,r}^{T} \in \mathcal{R}^3
% \end{split}
% \end{equation}

\begin{equation}\label{eqn:QR_Wb}
    W= \operatorname*{arg \space max}_{W} \sum_{r \in \Upsilon}\log (s_{c_r})
\end{equation}

% \[W^{T} = [w_1 \quad w_2 \quad ... \quad w_n] \qquad X_{i,r}^{T} \in \mathcal{R}^3
% \]
We use Maximum Likelihood Estimation as shown in \eqref{eqn:QR_Wb} to determine the values of strategic prioritization $W$ that best fit the data.
%this strategic prioritization, $W$.

% where $X$ represents a vector of values for attributes of the choices, and $W$ is the strategic prioritization showing how much weight a participant gives to each attribute.

\section{Experimental Setup and Protocol}\label{sec:experiment}

%The following sections describe our setup and procedure. %We also provide a Github link with code and instructions so anyone with the appropriate robot can replicate this study.\footnote{Code to be supplied after review.}  TODO include if code up

\subsection{Participants}\label{sec:exp_participants}

We recruited $40$ participants from the local community ($15$ M, $24$ F, $1$ nonbinary, $M_{age}=27.2, SD_{age}=11.2$). All participants played Game Session I consisting of 35 rounds of the Guards and Treasures game against the robot, referred to as the ``basic game''.  A selected group of the participants, referred to as the ``two-session group'', also played Game Session II, consisting of another 35 rounds, referred to as the ``additional games'', in which the robot exhibited the opposite language behavior as from the basic games.
%All participants played at least one game session against the robot (a game session consisting of thirty-five rounds of the game), the ``basic game''.   Some (the ``two-session group'') were selected to play an additional game session (``additional games'') in which the robot exhibited the opposite language behavior. 

\subsection{Robot}\label{sec:exp_robot}

We used the Pepper Robot by Softbank Robotics~\cite{guizzo2015robot} (a research robot % OPT-DEL this phrase
provided for participation in the RoboCup Social Standard Platform League).\footnote{http://www.robocupathome.org/athome-spl} Pepper is a humanoid robot with arms, a head with cameras and microphones, mobility, and voice abilities. See it pictured in Fig.~\ref{fig:study}.
HRI research suggests that physically present, embodied robots may make interactions more engaging and enjoyable and increase social presence~\cite{kiesler2008anthropomorphic, powers2007}. We used a humanoid robot for these reasons, as well as to give the participants a visual, physical reference for interacting with their opponent (for example, Pepper made eye contact with participant). %, and was also able to use gestures to remind the participant of its engagement. 
We hoped that using an embodied robot would maximize the opportunity for participants to attribute social characteristics to the opponent and exhibit task-relevant and affective responses to its behaviors. Pepper acted autonomously according to a script that was set before each game session.

\subsection{Procedure}\label{sec:exp_procedure} The experimental procedure was as follows:
\subsubsection{Consent} The experimenter obtained written consent to participate in the study and verbally informed the participant that video and audio recordings of the session would be made.
\subsubsection{Pre-Game Survey} Before the game, the experimenter administered a questionnaire to collect demographic information and measures of pre-task emotional state.% (more details about the pre- and post-game questionnaire are in section \ref{sec:exp_surveys}).
\subsubsection{Practice Rounds} The participant played on a ``convertible'' combination laptop/tablet. In order to counter learning effects, we had participants play two practice rounds of the game ``against the computer''.  %This countered learning effects in participants' performance, as they learn to play before encountering the robot.
\subsubsection{Game Session I (Basic Games)} After the practice rounds, the participant was led into a room where the robot sat behind a table. The participant sat across from the robot with the tablet %(the same as from the practice round, but now plugged in) 
face-up between them. % A video camera was set up behind the table.  
%While the participant played the game, experimenters sat in a different part of the room and were hidden from view by a screen to prevent the participant from feeling like they were being observed.
The participant then played several rounds of the game ``against'' the Pepper robot. The robot made periodic comments about the game and the participant. The comments exhibited either \textbf{encouraging} or \textbf{discouraging} expressive affect. % Each ``encouraging'' comment was paired with a ``discouraging'', and the comments in each pair only differed by one word.  The robot's comments were determined by  the Affect condition and, 
Although the commentary was sometimes complimentary and sometimes critical in nature, in reality it had nothing to do with the participant's actual performance.  
\subsubsection{Post-Game Survey and Video} Upon completion of the games, the participant notified the researcher that they had completed the task. The researcher then administered a written survey and a verbal semi-structured interview, which was video-recorded. 
\subsubsection{Game Session II (Additional Games)}\label{sec:gs2a} A selected subset (the ``two-session group'') of participants played a second game session against the robot, which exhibited the opposite affect as from the first game.
\subsubsection{Post-Game Survey and Video II}\label{sec:gs2b} If a participant played a second game session, they were given a second post-game survey and asked the same set of verbal questions again.
\subsubsection{Debriefing} Initially, participants were told they would play against a robot but were not informed of the true purpose of the study. Participants were debriefed after the study ended. 

\subsection{The Guards and Treasures Game Interface}\label{sec:exp_game}

\begin{figure}
    \centering
    \includegraphics[width=0.3\textwidth]{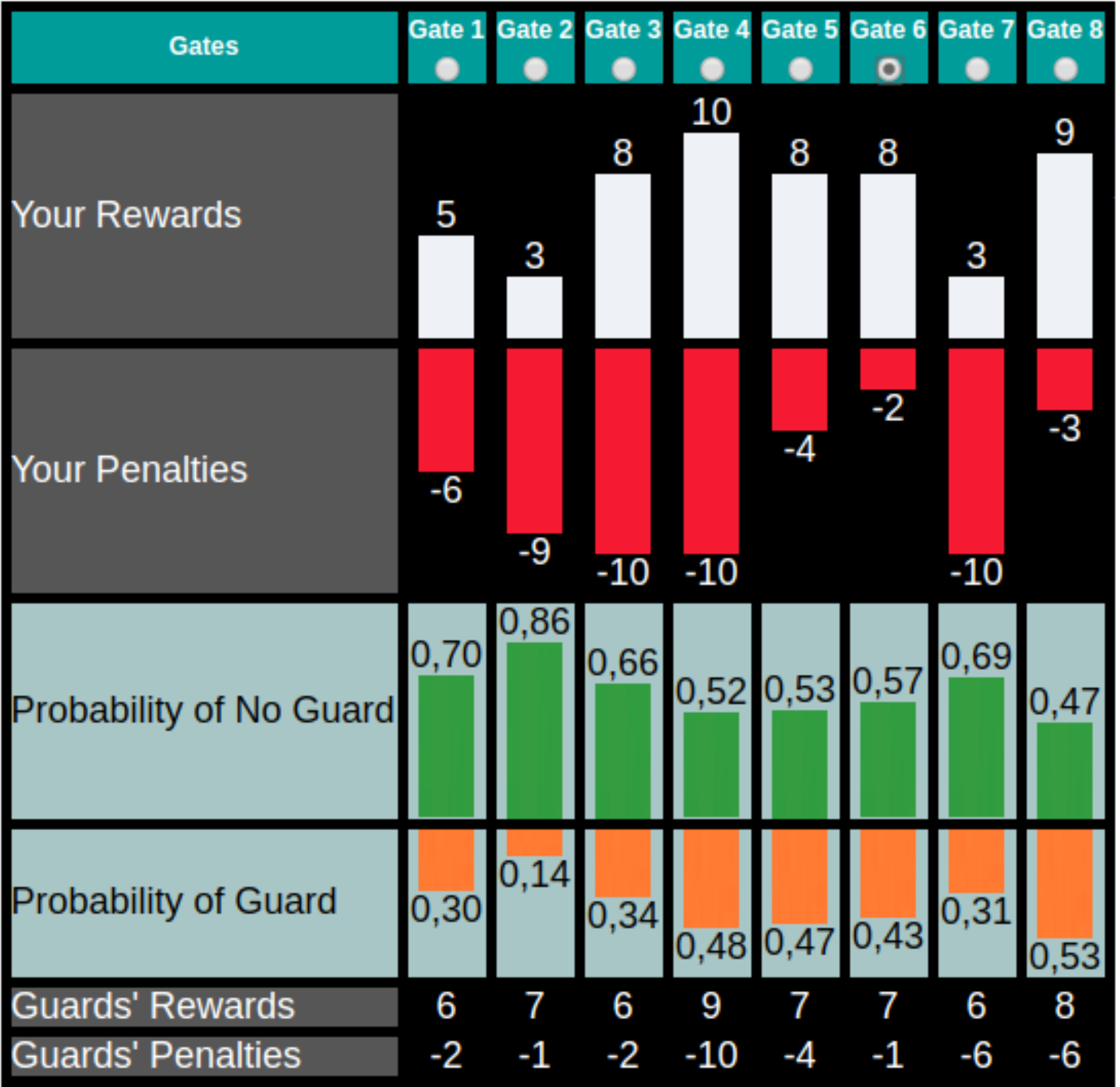}
    \caption{A round of the Guards and Treasures game, adapted from \cite{yang2011improving}.}
    \label{fig:gate_game}
\end{figure}

% Human behavior in the context of a Stackelberg Security game has been studied extensively in computational game theory.
One round of the ``Guards and Treasures'' game %\footnote{The code we used to run the game can be found at CODE and is open source.} 
%that the participants played consists of 35 rounds (following 2 practice rounds). It 
is a modified version of the game from \cite{yang2011improving}.\footnote{This original game can be found at \url{http://teamcore.usc.edu/Software.htm}.} This specific game is useful for studying bounded rationality. It features a virtual scenario and provides a limited set of options in each round.  In each round, the participant is shown a screen as in Fig.~\ref{fig:gate_game}. The central idea of the game is that the player can choose to ``attack'' (select) each of several gates. If the defending player (the robot) places a guard at the gate, the human player incurs the penalty for that gate. If the chosen gate is not guarded, the human player receives the reward instead. The probability that a guard is behind a particular gate is also displayed. The player selects one gate each round and only sees their results (whether the gate they chose is guarded and whether they got the reward or penalty) after all the rounds are complete.

The ``choices'' referred to in section \ref{sec:methods_QR} are the various gates.
The expected utility $U_{i,r}$ of a particular gate $i$ for a given round $r$ (referenced in \eqref{eqn:QR_Lb}) can be found by
\begin{equation}\label{eq:U}
U_{i,r} = R_{i,r} (1-g_{i,r}) - g_{i,r} Y_{i,r}
\end{equation}
where $R$ is the reward, $g$ is the probability a guard will defend a gate, and $Y$ is the penalty for a particular gate $i$ in round $r$.  ($R \in \mathbb{Z}, 
%1 \leq R \leq 10 $, 
R \in [1, 10],
Y \in \mathbb{Z},
% $-10  \leq Y \leq -1$, 
Y \in [1, 10]$,
and $g \in [0,1]$)  $N=8$ as there are 8 gates, each gate is one of the choices a participant can make. The $X_{i,r}$ from \eqref{eqn:QR_Wb} is $X_{i,r}^{T} = [R_{i,r} \quad Y_{i,r} \quad g_{i,r}]$.  $W$ has three components, $w_1$, $w_2$, $w_3$, which refer to how much weight a participant gives to reward, penalty, and probability of seeing a guard, respectively.

\begin{table*}[h]
\caption{$\lambda$ and $W$ for various populations $\Upsilon$ \label{tbl:QR}}
\begin{center}
\resizebox{1\textwidth}{!}{
\begin{tabular}{|l|c|c|c|c|c|c|c|c|c|c|c|c|c|c|c|}
\hline
& \multicolumn{3}{c|}{$\lambda$} & \multicolumn{9}{c|}{$W$}\\
\hline
\multicolumn{1}{|r|}{\textbf{Affect:\quad}} & \textbf{Both} & \textbf{Positive} & \textbf{Negative} &
\multicolumn{3}{c|}{\textbf{Both}} &
\multicolumn{3}{c|}{\textbf{Positive}} & \multicolumn{3}{c|}{\textbf{Negative}} \\
\hline
\textbf{Rounds $\Upsilon$ from Population:}&&&& $w_1$ & $w_2$ & $w_3$& $w_1$ & $w_2$ & $w_3$& $w_1$ & $w_2$ & $w_3$\\
\hline
Basic Games For All & 0.5432 & 0.5828 & 0.5064&0.3261 & 0.1697 & -10.4838 & 0.3586 & 0.1573 & -11.1006 & 0.2965 & 0.1819 & -9.939\\
\hline
Basic Games for Two-Session Group & 0.3269 & 0.2256 & 0.3929 &0.1649 & 0.1061 & -8.007& 0.0893 & 0.0818 & -5.6158 & 0.2190 & 0.1254 & -9.7572\\
\hline
Additional Games for Two-Session Group & 0.4128 & 0.4892 & 0.3015 &0.1907 & 0.2021 & -10.2328 & 0.0742 & 0.2028 & -9.7686 & 0.2624 & 0.2041 & -10.6888\\
\hline
Basic and Additional Games for All & 0.5121 & 0.5568 & 0.4660&0.2947 & 0.1761 & -10.3758& 0.3318 & 0.1692 & -10.9512 & 0.2564 & 0.1840 & -9.8081\\
\hline
\end{tabular} }
\label{tab1}
\end{center}
\end{table*}

\begin{figure*}[!b]
    \centering
    \mbox{
    \subfigure[Value of $\lambda_P$ (positive) and $\lambda_N$ (negative) over time (captured at seven 5-round intervals across an affect class) Both increase, but $\lambda_P$ more so \label{fig:lambda_trends}]{\includegraphics[height=3.8cm]{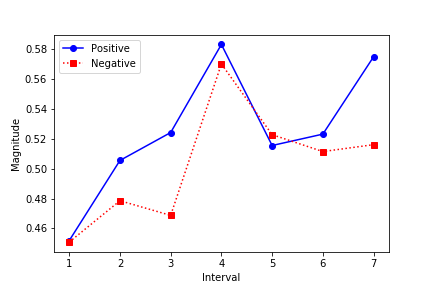}}\quad
    \subfigure[Values of $w_1$ and $w_2$ (Reward and Penalty components of $w$ over time. Each interval is 5 rounds.) Participants place more value on penalty over time. \label{fig:W_trends}]{\includegraphics[height=3.8cm]{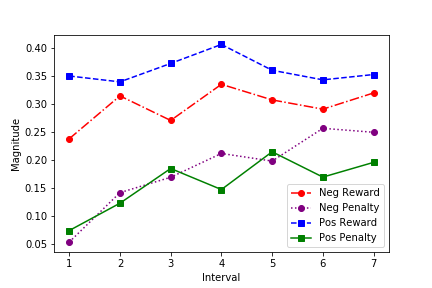}}\quad
    \subfigure[Mean scores for Self Assessment Manikin for perception of robot by affect class. Error bars represent $\pm 1$ standard error of the mean%. Robot behavior influenced people's feelings.
    .\label{fig:robotSAMbyaffect}]{\includegraphics[height=3.8cm]{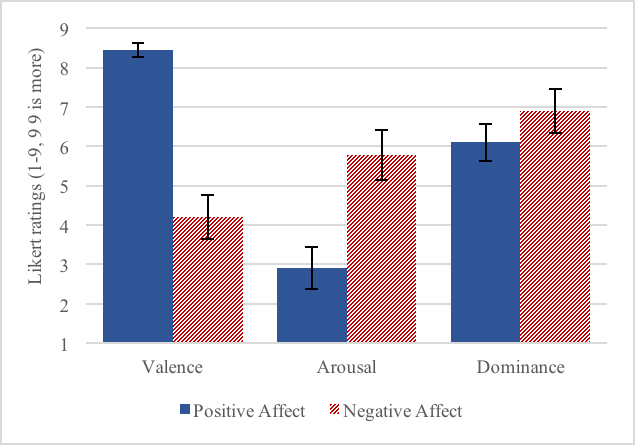}}
  }
    \caption{Results: Trends and Comparisons \newline
    }
\end{figure*}

\subsection{Surveys and Data Collection}\label{sec:exp_surveys}

Our measures consisted of i) a pre-task questionnaire, ii) records of actions taken during the game, iii) a post-task questionnaire, iv) a post-game verbal semi-structured interview (recorded on camera), and (for some participants) v) video of the participant playing the game against the robot.%\footnote{Exact pre- and post-surveys used are available for download at [Link to be added before publication].} %TODO 
%Participant data was anonymized, encrypted, and stored according to assigned IDs. 
%Participant data was anonymized and encrypted.

The pre-task questionnaire included demographic information and numerical ratings of familiarity with robots and with technology. The post-task questionnaire asked the participant to assess their own performance and their experience with the game and the robot.  %The aforementioned questions are on yes/no or numeric scales. Free-response questions related to a person's strategy and their perceptions of the robot were also asked. 
We originated some of the questions, and other questions we drew from \cite{mutlu2006storytelling} and \cite{suk2006color}. The pre-task and post-task questionnaires both made use of the Self Assessment Manikins\cite{bradley1994measuring}, which measure affect via three dimensions: valence, arousal, and dominance. In the first questionnaire, participants assessed themselves on these scales, and in the second, they assessed both themselves and the robot.

In a post-task semi-structured interview, we asked 9 questions pertaining to participants' overall perception of the robot, overall thoughts about the experience, self-assessment of performance, perceptions of the robot's goal, and game-playing strategy.
Participants who played the additional session answered the pre-task questionnaire once and the post-task questionnaire and interview questions after each game session.

\section{Results}\label{sec:results}

\subsection{Analysis on Gameplay}
We solve \eqref{eqn:QR_La} and \eqref{eqn:QR_Wb} over data from rounds from aggregated groups of participants. 
In table \ref{tbl:QR}, we show $\lambda$ and $W$ values corresponding to rounds played by various populations ($\Upsilon$s).  These $\Upsilon$s 
%contain
span
multiple participants, divided by affect, and include all rounds for each participant in each subset. In Quantal Response mode, $\lambda = 0$ indicates uniform random behavior and $\lambda = \infty$ indicates perfect rationality (i.e., best response). % Find the same split for $W$ values (the strategic prioritization, our variant of the Subjective Utility Quantal Response) in table \ref{tbl:W}, where $W$ describes a participant's strategy prioritization ($w_1$ is the reward component, $w_2$ is the penalty component, and $w_3$ is the probability of guard component).
% $W$ values (the strategic prioritization, our variant of the Subjective Utility Quantal Response) describes a participant's strategy prioritization ($w_1$ is the reward component, $w_2$ is the penalty component, and $w_3$ is the probability of guard component).
The parameter values of $W$ in our variant of the SUQR model describes a participant's strategic prioritization over different factors influencing their decision making ($w_1$: reward component, $w_2$: penalty component, $w_3$: probability-of-guard component).
For reference, previous work ~\cite{nguyen2013analyzing} obtained $\lambda = 0.77$ via a group of Amazon Mechanical Turk (AMT) workers playing this game online.
In addition, ~\cite{nguyen2013analyzing} reports $W = [0.37, 0.15, -9.85]$ (converted to our representation) for the SUQR model for a general population playing this game.

The first row in the table has basic, first round games for all participants. The ``two-session group'' of participants are those who played an additional game with the reverse affect (mentioned in sections \ref{sec:gs2a} and \ref{sec:gs2b}). 
In addition, using the procedure in \cite{Pita}, we analyze changes in $\lambda$ and $W$ between the basic and additional session for each individual participant in the two-session group. We found that those who played a negative session first had a 21\% increase in $\lambda$ and a 104\% increase in the 1-norm of the strategic prioritization vector, $W$, on average.
On the other hand, 
those who played the positive session first had a 28\% increase in  $\lambda$ and a 110\% increase in the 1-norm of $W$. 
Further, we divide the 35 rounds of game into seven 5-round intervals and analyze the best parameter values in each stage.
%(These percentages represent averages of percentage change per-participant, as opposed to the table which has $\lambda$ and $W$ calculated across rounds aggregated from populations.)
Fig.~\ref{fig:lambda_trends} shows the trend for $\lambda$ for positive and negative affect over time for participants' basic games.
%DEL-OPT: 2 lines above, get rid of 'on the other hand'

We notice that although participants place a higher priority on reward than on penalty, all participants place more emphasis on penalty over time in Fig.~\ref{fig:W_trends}. Just as weight of reward ($w_1$) is relatively steady over multiple intervals of five rounds, we found that the weight participants placed on the probability of a guard being present ($w_3$) was steady over time, though the weight placed on the guard's presence ($w_3$) was two orders of magnitude larger than the other two weights.
In the positive affect condition, 15 participants won the game (i.e., the total utility the participant achieved was higher than their opponent) and 5 lost. In the negative affect condition, 16 participants won and 4 lost. This difference in performance was not significant. 
% We also did not find any effects of robot affect on
% We found no effect of robot affect on
We found no effect of affect on prioritization between W components.
% prioritization of the guard, reward, and penalty factors. 
%Need to make sure the GRP things are mentioned in methods and/or measures

%\subsection{Perception of Robot}

%Age was relevant in participants' perceptions of the robot. Negative expressive language led participants to have a greater perception of Pepper having human-like traits than younger participants.are more used to thinking of robots as machines.

\begin{figure*}[h!]
\begin{center}
\resizebox{0.9\textwidth}{!}{
\begin{tabular}[c]{l|l}%\label{tbl:ngram_sentences}
\textbf{Positive Sentences} & \textbf{Negative Sentences} \\
\hline
%What a wonderful choice you made there. & What a awful choice you made there. \\
You seem to be considering your moves in a practiced manner. & You seem to be considering your moves in a bizarre manner. \\
Honestly this game is a wonderful experience. & Honestly this game is a bad experience. \\
I have to say you are a great player. &
I have to say you are a terrible player. \\
Over the course of the game your playing has become brilliant. & Over the course of the game your playing has become confused.
\end{tabular}
}
\end{center}
\caption{Example sentences built by our affective NLP algorithm}
\label{fig:ngram_sentences}
\end{figure*}

\subsection{Analysis of Self Assessment Manikin}

The questionnaire about perceptions of the robot was also analyzed. %A \textit{t}-test was run to assess the residuals of the outcome variables. 
Because the data were not normally distributed and the sample was small (\textit{n} $< 50$), the non-parametric Wilcoxon/Kruskal-Wallis rank sum test for two samples was used to compare the two conditions for all variables.

The robot's expressive language significantly affected several measures of positive social assessments of the robot, including perceptions that the robot was encouraging, $\chi^2(1, N = 40) = 31.55$, $p = 0.008$, optimistic, $\chi^2(1, N = 40) = 23.48$, $p < 0.0001$, and cheerful, $\chi^2(1, N = 40) = 28.33$, $p < 0.0001$. For all of these variables, encouraging language led to higher ratings. There was no effect of the language manipulation on perceptions of humanlikeness or cuteness. These results serve as a validity check, suggesting that the encouraging commentary was perceived as positive and   and the discouraging commentary was perceived as negative.

We found a significant main effect of robot affect on participants' liking of the activity, $\chi^2$(1, \textit{N} = 40) = 6.97, \textit{p} = 0.008. We used the Self Assessment Manikin (SAM) scale \cite{bradley1994measuring} to measure the participants' mood and before and after the game in terms of emotional valence (how happy or unhappy they felt), arousal (how excited or unexcited they felt), and dominance (how in-control they felt). We also used this scale to assess participants' perceptions of the robot's mood after the game. There was a significant main effect of robot affect on post-task participant valence, $\chi^2$(1, \textit{N} = 40) = 4.36, \textit{p} = 0.037, and perceived robot valence, $\chi^2$(1, \textit{N} = 40) = 20.87, \textit{p} $<$ 0.0001.%, Fig.~\ref{fig:robotSAMbyaffect}.  
For these variables, participants in the encouraging language condition had higher ratings. We also found an effect wherein discouraging language positively impacted ratings of perceived robot arousal, $\chi^2$(1, \textit{N} = 40) = 10.07, \textit{p} = 0.002. There were no main effects of language on post-task participant arousal, participant dominance, or perceived robot dominance. 
The effect of condition on perceptions of robot affect can be seen in Fig.~\ref{fig:robotSAMbyaffect}. %(see Figure~\ref{fig:affectSAMscores}).
Collectively, these findings corroborate previous work \cite{anger,cooperate} suggesting that affective robot behavior is able to strongly influence people's feelings in a dyadic interaction. Our case differs from this work in that the setting is competitive rather than cooperative.

We suspected that other independent variables such as age, gender, preconceived notions about robots, and mood prior to the experiment may also play a role in evaluations of the robot. We looked for correlations among these variables and ran our analyses again with correlated variables as covariates. We found a main effect of age on participants' belief that the robot was humanlike, \textit{p} = 0.003, in which younger participants thought it was more humanlike. We also found a significant interaction effect of age and expressive language condition on perceptions that the robot was humanlike, \textit{p} = $0.012$, and ratings of the robot's \textit{dominance}, \textit{p} = $0.004$: negative affect mattered less for younger participants in assessments of humanlikeness and dominance. We found an interaction effect of pre-task participant valence and robot expressive language on perceptions that the robot was humanlike, \textit{p} = 0.015, in that discouraging language and low valence prior to the start of the experiment led to lower perceptions of humanlikeness. 

We found that discouraging language significantly lowered participants' beliefs that the robot was optimistic, \textit{F}(1, 9) = 65.05, \textit{p} $<$ 0.0001, cheerful, \textit{F}(1, 9) = 45.64, \textit{p} $<$ 0.0001, and cooperative, \textit{F}(1, 9) = 24.77, \textit{p} = 0.008. This was similar to the findings from our between-subjects analysis. Here, we also found that encouraging language increased perceptions that the robot was cute, \textit{F}(1, 9) = 6.92, \textit{p} = 0.027.

\subsection{Participant Interviews}\label{sec:results_oral}
To gain further insight into participants' impressions of the robot, we conducted semi-structured interviews with our participants. Twelve participants (four in the encouraging condition and eight in the discouraging condition) reported a belief that the robot's goal involved distracting them. Participants in the encouraging condition said, ``When I was trying to determine what move to make, it took me out of that zone for a bit'' (P220), and, ``It felt like I was doing homework and my friend kept talking to me'' (P201).  Altogether, 30\% of participants explicitly classified the robot's goal as ``distraction''. Participants also spoke about the robot's behavior as a result of its programming. For example, P104 said, ``I don't like some of the stuff it was saying.  But that's the way it was programmed so I can't blame it''. Interviews also further confirmed participants were encouraged by the robot in the encouraging condition and were especially discouraged in the discouraging condition. When asked about the robot's goal, a participant in the encouraging condition answered, ``To encourage me to do well... it seemed to [succeed in that goal]'' (P117), while a participant exposed to the discouraging language said ``It kept making me doubt myself'' (P214).

\section{Discussion}\label{sec:disc}

\subsection{Validation of NLP Model}\label{sec:disc_nlp}
Participants perceived an encouraging robot as encouraging, cheerful, and optimistic, and a discouraging robot as discouraging and pessimistic. Interviews supported the quantitative results.
% (As mentioned, the code is available for other researchers to use in their studies.) 
Example sentences generated by the model can be found in Fig.~\ref{fig:ngram_sentences}.
This validates the affect-aware bidirectional fill-in-the-blank N-gram NLP model we developed, and demonstrates that our simple word choice model achieved the desired result. 

\subsection{Population Rationality}

Overall, we found that discouraging expressive language caused less-rational performance ($\lambda = 0.51$ for negative vs. $\lambda = 0.58$ for positive). This is in line with what might be expected and with previous work \cite{nguyen2013analyzing,ilah,tidy}. A participant will believe they will make better choices when encouraged, whereas a discouraged individual will make more mistakes.  %
% find a source to prove this  -- NOTE: you can include a human-encouraging-a-human as a source
Pepper's form is particularly similar to that of a human (two arms, fingers, head, torso), so certain aspects of the interaction may more closely mimic human-human game play than they would have if our participants had played with a less humanoid robot.

Participants who played an additional game (players in the ``two-session group'') performed more rationally and more strategically (as noted by the increase in the 1-norm of $W$) in the additional session compared to the basic session.  Those who played the positive affect session first had a higher increase in these metrics than those who played a negative affect session first.  One possible explanation is that in the first case, residual encouragement from the initial positive session continued to buoy the participant in the second session.

While there were outliers, % individuals,
our participants' overall rationality was below that of the crowd-sourced AMT population from~\cite{nguyen2013analyzing}. %Amazon Mechanical Turk (AMT) population from ($\lambda = 0.77$) \cite{nguyen2013analyzing}.  
The discrepancy may be attributable to differences in the game framing, timing, population, or noise. 
One possible explanation is that the amount of money our participants received was fixed as opposed to dependent on their performance, like AMT workers. 
%In other words, they themselves derived no benefit from how much effort they actually put in. 
The other major difference between that study and our own is that AMT workers were competing in the game against a computer, while physically located in (presumably) a location of their choice.  Our participants were face-to-face with a robot ``opponent'' in an unfamiliar room.
An unfamiliar setting can influence a participant's decision rationale and may have been an additional factor hampering the competitive abilities of many participants~\cite{nooraie2012factors}. Dialogue can also be a distraction, regardless of content~\cite{drews2008passenger}.

%Another reason for decreased rationality may be the competitive nature of the task.  While emotion is contagious in a cooperative setting (robot encouragement would be expected to help a human), that may not apply in a competitive setting%~\cite{gilbert2017measuring}.  
Over a quarter of all participants \textbf{explicitly} expressed a belief that that one of the robot's goals was to distract them.  This suggests that, given the competitive setting, some participants were focused more on winning the game than on interacting with the robot. Another reason for decreased rationality may be the competitive nature of the task.  While emotion is contagious in a cooperative setting (robot encouragement would be expected to help a human), it may not be in a competitive setting.

\subsection{Perception of Humanoid Robot}
While many individuals anthropomorphized the robot, multiple participants described the robot in ways that dehumanized it. This awareness or assumption of the robot's lack of agency (despite the fact that it was autonomous) could also have contributed to a participants being less impacted by it overall.
%Negative expressive language led participants to have a smaller perception of Pepper having human-like traits than younger participants, 
Younger participants were less influenced by affect, which could be due to a younger generation more used to thinking of robots as machines.

\section{Conclusion}\label{sec:conclusion}

A humanoid robot that encourages or discourages a human opponent can impact that human's rationality. In our study, a discouraging robot led to lower rationality while an encouraging robot was associated with higher rationality.  % TODO consider adding: Our study can be replicated by using the open source code we provide along with the appropriate hardware.
The insights documented here may be useful for future designers of robots.  Game developers can also use this knowledge to create more interactive opponents to increase the sense of engagement and enjoyment.
In the field of education, we can be aware that were a humanoid robot exam proctor to express affect in its language while administering an exam to students, the students' performance could be influenced, for better or for worse.
Our findings may serve to help future robot designers develop a better understanding of how affect impacts perceptions of a social robot during non-cooperative interactions.
%We have developed open source tools, such as the program to generate and run the affect-aware bidirectional fill-in-the-blank N-gram NLP model, as well as code for running it and the whole experiment on the Pepper robot.  These resources are validated as serving their intended purpose and we invite other researchers to take advantage of and use them.
Useful future work would be to investigate nonverbal modes of expression, like body movement and gestures, in competitive settings. % It would also be worthwhile to investigate whether rationality and other measures discussed here are influenced in a similar manner in additional types of competitive settings.

% % use section* for acknowledgment
\section*{Acknowledgment}

This work was supported in part by NSF grant IIS-1850477.
We would like to thank Jeffery Cohn (Robotics Institute/Department of Psychology, CMU/Pitt) and Louis Philippe Morency (Language Technologies Institute, CMU) for their guidance and advice. We thank Tianyu Gu and Ashley Liu for help running some of the experiments.  We thank Gayatri Shandar who transcribed and annotated some of the videos.



\balance
\bibliographystyle{IEEEtran}
\bibliography{bibtex/refs.bib}
\nocite{roth2018impact}


\end{document}